\documentstyle[12pt,aasms4,epsf]{article}
\received{00 December  1997}
\accepted{00 1996}
\centerline{Submitted to the Editor of the ApJ {\it 
Letters} on December  7, 1997}
\lefthead{Yusef-Zadeh, Roberts \& Biretta}
\righthead{High Velocity Ionized Gas}

\def\deg    {$^{\circ}$}                        
\def\kms    {\hbox{km{\hskip0.1em}s$^{-1}$}}    
\def\msol   {\hbox{$M_\odot$}}                  
\def\dasec  {\hbox{$.\!\!^{\prime\prime}$}}     
\def\asec   {$^{\prime\prime}$}                 
\def\dasec  {\hbox{$.\!\!^{\prime\prime}$}}     
\def\dsec   {\hbox{$.\!\!^{\rm s}$}}            
\def\min    {$^{\rm m}$}                        
\def\hour   {$^{\rm h}$}                        
\def\amin   {$^{\prime}$}                       

\begin{document}

\title{Proper Motions of Ionized Gas at the Galactic  \\
Center: Evidence for Unbound Orbiting Gas}

\author{F. Yusef-Zadeh}
\affil{Department of Physics and Astronomy, Northwestern University, 
Evanston, IL 60208 (zadeh@nwu.edu)}

\author{D. A. Roberts}
\affil{NCSA, 1002 W. Green St., Urbana, IL 61801
 (dougr@ncsa.uiuc.edu)}

\author{J. Biretta}
\affil{STScI, 3700 San Martin Drive, Baltimore, MD 21218
(biretta@stsci.edu)}

\begin{abstract}

We present radio continuum observations of the spiral-shaped ionized
feature (Sgr A West) within the inner pc of the Galactic center at
three epochs spanning 1986 to 1995.  The VLA A-configuration was used
at $\lambda$2cm (resolution of 0\dasec1$\times$0\dasec2).  We detect
proper motions of a number of features in the Northern and Eastern
Arms of Sgr A West including the ionized gas associated with IRS 13
with V(RA)= 113$\pm$10, V(Dec)=150$\pm$15 km s$^{-1}$, IRS 2 with
V(RA)= 122$\pm$11, V(Dec)=24$\pm$34 km s$^{-1}$ and the Norther Arm 
V(RA)= 126$\pm$30, V(Dec)=--207$\pm$58 km s$^{-1}$.  We also report the
detection of features having transverse velocities $>$1000 \kms\
including a head-tail radio structure, the ``Bullet'', $\approx$4$''$
northwest of Sgr A$^*$ with V(RA)= 722$\pm$156, V(Dec)=832$\pm$203 km
s$^{-1}$, exceeding the escape velocity at the Galactic center.

The proper motion measurements when combined with previous H92$\alpha$
radio recombination line data suggest an unambiguous direction of the
flow of ionized gas orbiting the Galactic center.  The measured
velocity distribution suggests that the ionized gas in the Northern
Arm is not bound to the Galactic center assuming a 2.5 million solar
mass of dark matter residing at the Galactic center.  This implies
that the stellar and ionized gas systems are not dynamically coupled,
thus, supporting a picture in which the gas features in the Northern
Arm and its extensions are the result of an energetic phenomenon that
has externally driven a cloud of gas cloud into the Galactic center.

\end{abstract}

\keywords{galaxies:  ISM---Galaxy: center ---ISM: individual 
(Sgr A East and Sgr A West) --- ISM: magnetic fields}

\vfill\eject

\section{Introduction}

The ionized gas known as Sagittarius A West (Sgr A West) appears as a
three-Arm spiral-like structure (North, East, and West Arms) engulfing
the inner pc of the Galaxy where Sgr A$^*$, the compact radio source
at or near the dynamical center of the Galaxy lies (Ekers et
al. 1983). These features are surrounded by neutral gas in the
circumnuclear disk (CND) rotating with the velocity of about 100 \kms\
at the distance of 2 pc from the Galactic center (e.g. G\"usten et al.
1988).  The kinematics of ionized gas surrounding Sgr A$^*$ show
systematic velocities along various components of Sgr A West including
Western Arc
with a radial velocity
structure which varies regularly between --100 and +100 \kms\ in the
North-South direction (e.g. Serabyn et al. 1988; 
 Herbst et al. 1993; Roberts \& Goss 1993). 
  However, the velocity
structure of the inner 10$''$ where there is a hole in the
distribution of ionized gas, known as the ``mini-cavity'', becomes
increasing more negative $\approx-$350 \kms\ approaching Sgr A$^*$
(Yusef-Zadeh, Morris \& Ekers 1989; Roberts, Yusef-Zadeh \& Goss 1996,
hereafter RYG).

Recent observations of stellar proper motions shows evidence of a
2.5$\times10^6$ \msol\ object lying close to the position of Sgr
A$^*$ (Eckart and Genzel 1997).  The stars orbit randomly around the Galactic center with
increasing velocity dispersion around Sgr A$^*$, reflecting the
gravitational potential of central mass.  The ionized gas, on the
other hand, is part of a coherent flow with systematic motion that is
decoupled from the stellar orbits. Understanding the kinematics of the
system of ionized gas is complicated by its incomplete view of the
3-dimensional geometry with respect to Sgr A$^*$ as well as by the
interaction of orbiting gas with non-gravitational forces, such as the
winds from the cluster of hot mass-losing stars near the Galactic
center. To examine the gas kinematics and the true geometry of the
ionized flow, in this {\it Letter} we present the results of proper
motion measurements of ionized gas at $\lambda$2cm.


\section{Observations}

Radio continuum observations were carried out with the Very Large
Array of the National Radio Astronomy Observatory\footnote{The
National Radio Astronomy Observatory is a facility of the National
Science Foundation, operated under a cooperative agreement by
Associated Universities, Inc.}  in 1986.167, 1990.375 and 1995.557 at
$\lambda$2cm in the A-configuration using similar {\it uv} coverage,
identical bandwidths of 12.5 MHz and identical phase center
at RA(1950)=17\hour 42\min 29\dsec33, Dec(1950)=$-29$\deg 59\amin 17\dasec0.
  We have employed a technique in
which the variability of Sgr A$^*$ is removed on every ten-minute scan
to improve the dynamic range of radio continuum images.  In order to
minimize the difference in calibration errors between three different
data sets, they were cross calibrated 
against the 1990.375 epoch before the 
final images were constructed.
To measure the two-dimensional transverse velocities with respect
to Sgr A$^*$, we generated Maximum Entropy images, and then applied
cross-correlation method of Biretta, Owen, \& Cornwell (1989)
  This method  which has been exhaustively 
tested (Biretta et al. 1995) determines the fractional 
pixel position shift between 
images (pixel size 35$\times$35 milli-arcseconds) 
which maximizes their cross-correlation.  Uncertainties on the velocities
were estimated by extracting 20 noise images from a CLEAN image in regions
beyond the inner 60\asec\ of the Galactic center, adding these to the MEM
images, and repeating the cross correlations; position uncertainties, and
hence velocity uncertainties, were determined from
the dispersion in the results for these 20 noise images.
 The rms noises of 1986,
1990 and 1995 images are 0.13, 0.104 and 0.16 mJy beam$^{-1}$,
respectively.


Sgr A$^*$ serves as an excellent reference point because of its small
proper motion: V(RA)=--13$\pm$13, V(Dec)=--11$\pm$9 \kms\ (Backer
1996), and is likely to be directly associated with the compact,
massive object.  The above technique was applied to two bright radio
continuum sources surrounding IRS 13 and IRS 10 continuum sources
The uncertainties
from our data for the east-west and north-south components of motion
are 11 and 33 \kms, respectively.  At the distance of 8 kpc, a shift
of 1 mas yr$^{-1}$ corresponds to about 39 \kms.  The velocity
uncertainty in the radial direction is much smaller than in transverse
directions.

\section{Results}  

Figure 1a shows the pseudo-color representation of the continuum image
at the resolution of 0\dasec1$\times$0\dasec2 with 22 boxes
superposed.  Boxes are shown where proper motion of ionized gas is
detected above 3$\sigma$ level, except Box 21, in at least one
direction ($\alpha$ or $\delta$).  Boxes 21 and 22 which surround IRS
1 and IRS 8, respectively, show transverse velocities at $< 3\sigma$
level, but nonetheless they are important because of the low upper
limits to their velocities.  Table 1 shows the transverse motions of
ionized gas averaged over the size of each box (column 5) in $\alpha$
and $\delta$ (columns 6 and 7).  The radial velocities averaged over
each box based on radio recombination line measurements (RYG) is shown
in column 8. The radial velocities of the features whose continuum
emission is too weak for H92$\alpha$ line detection or too compact are
not reported.  The angular distance of the center of each box with
respect to Sgr A$^*$ is shown in radial, $\alpha$, and $\delta$
directions in columns 2, 3 and 4, respectively.  The total measured
velocities and the upper limits to the corresponding escape velocities
of ionized gas (determined using the projected distance (r) from Sgr
A$^*$ with a mass of 2.5$\times10^6$ \msol) are shown in the last two
columns.

Figure 1b shows contours of $\lambda$2cm continuum image from A-array
observations at 0\dasec1$\times$0\dasec2 resolution superposed on
the pseudo-color representation of the radial velocity distribution of
H92$\alpha$ recombination line emission with spectral and spatial
resolutions of 14 \kms\ and 0\dasec75$\times$1\dasec2 (RYG),
respectively.  The vectors drawn from the center of each box show the
orientation of the transverse velocities and their lengths represent
their magnitude.  The regions with the highest signal-to-noise ratio
($>$10) averaged roughly over 1.5 square arcseconds are shown as Box 6
and 10 in Table 1 where the continuum emission is strongest.  The plots
in Figure 2 show the change in $\alpha$ and $\delta$ of the IRS 13 (Box 6)
source as a function of three epochs. Proper
motions were determined by a least-squared fit to the data where the
position values were weighted by their uncertainties.

The western half of the mini-cavity is represented by Boxes 6, 8, 10 and 11 
where the prominent clusters of hot stars, IRS 13 and IRS 2 are
embedded within extended ionized gas.  From [NeII] and H92$\alpha$
line observations the ionized gas in this region shows high radial
velocities $\approx -300$ \kms.  The transverse velocities vary from 24 and 150
\kms, which are much less than their radial velocity components.  The
filamentary ionized structure connecting the IRS 13 and IRS 2 (Box 11)
 features
also shows transverse velocities directed toward northwest.

The eastern half of the mini-cavity is represented by Boxes 1, 2 and 5.
Box 4 corresponds to the radio Blob $\zeta$ (Yusef-Zadeh, Morris \&
Ekers 1990; Zhao et al. 1991) lying near the center of the mini-cavity.
Again, all the features to the east and near the center of the
mini-cavity appear to have transverse velocities predominantly directed
to the west with the exception of the anomalous velocity feature in
Box 1 with transverse velocities exceeding 1000 \kms.  The trend of
westerly transverse motion of the ionized gas is also seen along the
Northern and Eastern Arms (Box 3, 12, 15, 16, 17, 18, 19).

Figure 1b also displays an anomalous high-velocity feature in Sgr A
West. This source called the ``Bullet'', as seen in Box 9, lies about
4$''$ northwest of Sgr A$^*$ and shows a total transverse velocity of
1100 \kms\ which is much greater than the upper limit to the escape
velocity of 373 \kms\ at its projected distance from Sgr A$^*$ (see
Table 1).  The low-resolution H92$\alpha$ measurements of this region
(Roberts \& Goss 1993) obtained using a limited velocity coverage
(--200 \kms\ $< V_{\rm LSR} <$ +200 \kms) show extended line emission
with radial velocities of about --36 \kms\ which is not likely
associated with the Bullet.  The peak flux density at $\lambda$1.2cm
is $\approx$2.5 mJy within a beam of 0\dasec31$\times$0\dasec22.
Assuming that the Bullet is an optically thin thermal source at the
temperature of 10$^4$ K, the electron density would be 3.6$\times10^4$
cm$^{-3}$, which is similar to typical electron densities of Blobs
(Wardle \& Yusef-Zadeh 1991).  The mass of the Bullet would be
8$\times10^{-4}$ \msol\ assuming a spherical size of radius
6$\times10^{-3}$ pc.
  
The proper motion velocity of the Bullet derived from the image
correlation technique is large enough that the positions of the peak
are displaced by a significant fraction of the synthesized beam on the
plane of the sky between the various epochs.  The velocity determined
from measuring the displacement of the peak agrees with that determined
by the
correlation method.  Figure 3 shows the transverse motion of the peak
emission in the three different epochs superposed on a greyscale
image of the Bullet (including the the tail) from the 1990
observations.  The most recent image (1995) shows the peak to be at
the position of the head of the source, with the tail pointing
roughly toward the head in the earlier observations.
The images shown in Figure 3 provide confidence in the results
obtained from two completely different techniques of proper motion
measurements.

\section{Discussion}

By combining the transverse and radial velocities, we are able to
unambiguously determine the direction of ionized flow at the Galactic
center.  The predominant component of the motion in the plane of the
sky is from east to west for most of the measured features with the
exception of few places where the velocity of ionized gas is
anomalously large.  It appears that the flow of ionized gas in the
Northern Arm (Box 12) originates in the northeast with red-shifted velocities
in the orbital plane.
The ionized gas then follows an orbital trajectory to the southwest as
it crosses the plane of the sky and passes by  Sgr A$^*$ before the
ionized gas moves to the northwest.  


One idea that has been suggested to explain the origin of the
mini-cavity and its unusual characteristics (e.g. Lutz et al. 1993; Melia et 
al. 1996) is the the collision of
fast-moving Blobs with the orbiting ionized gas. The Blobs are
hypothesized to be formed as a result of high velocity winds of IRS 16
cluster escaping from but focussed by the gravitational potential of
Sgr A$^*$.  This focusing mechanism allows the diffuse outflowing
materials to collide with each other and form dense Blobs of ionized
gas leaving the gravitational potential of Sgr A$^*$ (Wardle \&
Yusef-Zadeh 1992).  The anomalous high-velocity features seen in Boxes
1, 4, 8 and 9 are consistent with the outflow picture. In particular, the
Bullet is clearly escaping from the gravitational potential of the
Galactic center region even when the mass of the stellar cluster 
is included. In this model, however, it is not expected to see a
tail produced behind the fast moving Blobs.  The existence of a bow
shock or X-ray emitting gas associated with the head of the Bullet
would favor a model in which these fast-moving features are ejected by
mass-losing stellar sources.  Further observations of the Bullet are
needed to understand its origin.

The comparison between the measured total velocities and the upper
limits to the escape velocities at projected distance (r) from the
center of each box to Sgr A$^*$ (Table 1) suggests that the ionized
flow is on an unbound orbit around the Galactic center. This is
consistent with the model of RYG that the ionized gas in the Northern
Arm is on a hyperbolic orbit.  Like the compact and relatively 
isolated features with high transverse velocities discussed above (e.g. 
the Bullet and the Blobs), the rest of the 
orbiting ionized features are extended and follow a global velocity field
which may also be unbound to the Galactic center. 
If we use the projected distance as
cos$^{-1}$ (45$^\circ$) of the actual distance, almost all the measured
velocities are greater than the escape velocities listed in Table 1.
The mass within the inner 10-20$''$ is assumed to be dominated by a
compact source centered on Sgr A$^*$ having a mass of 2.5$\times10^6$
\msol\ as measured recently from stellar proper motion measurements
(Eckart \& Genzel 1997).  The escape velocity estimates may not be
applicable at large distances where the mass of the evolved stellar
cluster becomes important. From the comparison of the three
dimensional stellar and ionized gas motions, it appears that these two
systems are not dynamically coupled in the inner 20$''$ of the
Galactic center. 

The effect of a strong gravitational potential due to the large
concentration of dark matter near Sgr A$^*$ is manifested as high
velocity gradients of over 600 \kms\ pc$^{-1}$. However, the existence
of ionized gas in an unbound orbit is inconsistent with the notion
that the ionized gas in the Northern Arm is a tidally stretched
infalling feature (e.g. Serabyn et al. 1988).  Additionally, the
present proper motion data do not support the interpretation that the
Northern Arm is a segment of a one-armed spiral pattern induced as a
result of an instability in the rotating disk (e.g. Lacy et al. 1991).
A significant variation in the velocity distribution of ionized gas
along Northern Arm is not consistent with a small variations expected
from Keplerian motion, thus supporting that the Northern Arm is a
material feature.  We believe that the high velocity of ionized gas on
an unbound orbit supports a scenario in which an energetic phenomenon
outside the inner few parsecs of the Galactic center accelerated a
cloud to pass by the Galactic center, which then collides with the CND
and results in the loss of angular momentum of the material in the CND
(Serabyn et al. 1988).  There is evidence of disturbed neutral gas and
shocked molecular gas based on OH 1720 MHz maser emission at the
interface of the extension of the Northern and Eastern Arms of Sgr A
West and a ``gap'' in the CND (Yusef-Zadeh et al. 1996; Jackson et
al. 1993).  In this picture, the Northern and Eastern Arms delineate
the edges of the intruding cloud photoionized by the UV radiation
field at the Galactic center (Jackson et al. 1993).  The neutral gas
in the ``gap'' of the CND is interpreted to be the site of collision
with a cloudlet pushed into the Galactic center, possibly by the
energetic explosion of Sgr A East.  Future modeling of the
three-dimensional motion of ionized gas should constrain the
inclination of the orbital plane of the ionized gas with respect to
the orbital plane of the CND and examine the possibility that the CND
is origin of the Northern and Eastern Arms.

\acknowledgments

F. Yusef-Zadeh's work was supported in part by NASA grant NAGW-2518.
D. Roberts acknowledges support from the NSF grant AST94-19227. We 
thank Mark Wardle for useful discussion.

\begin{figure}
\plotfiddle{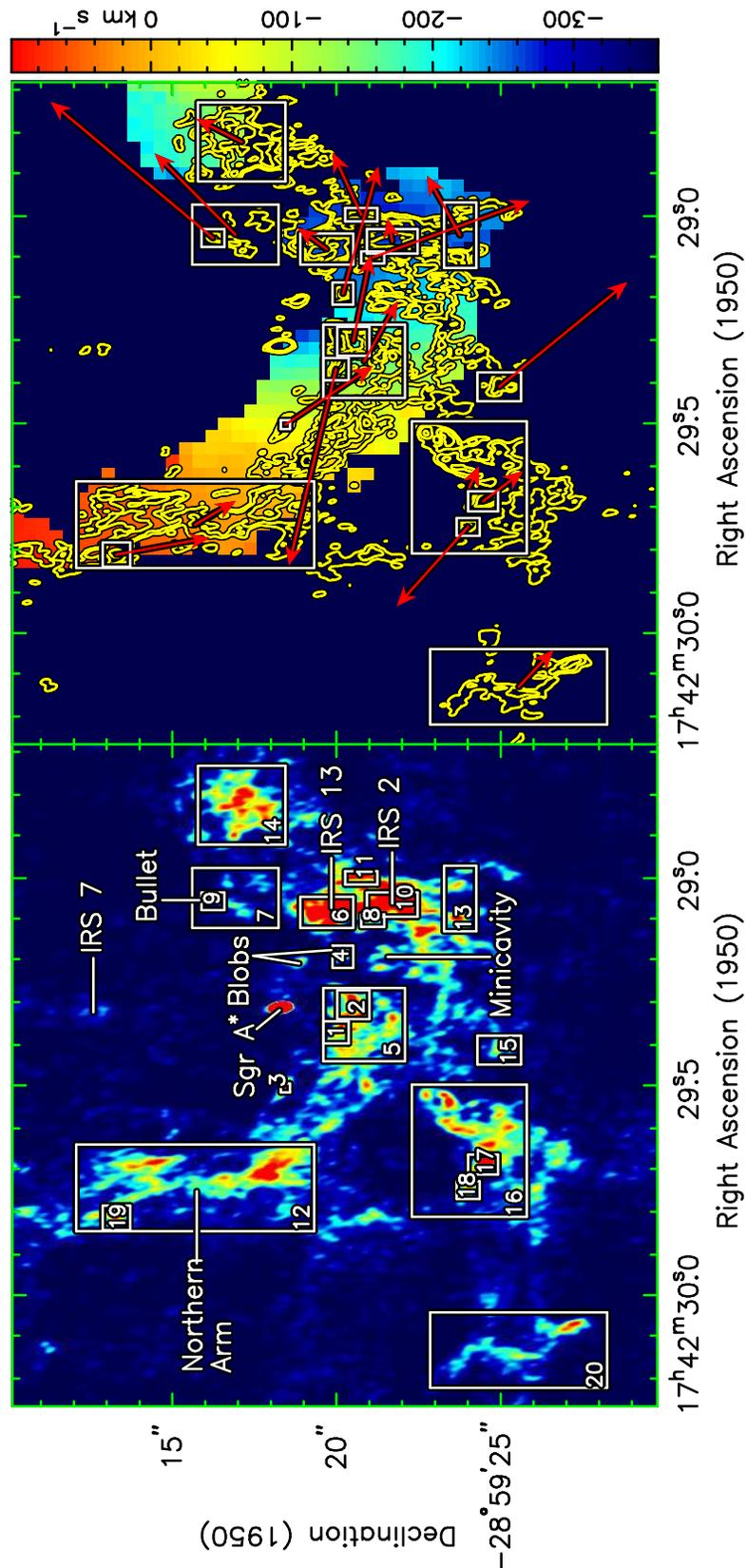}{6.7in}{0}{90}{87}{-250}{-40}
\figcaption{{\singlespace\small (a) A pseudo-color image of the
$\lambda$2cm continuum 
from the 1990 epoch data ({\it left}) with a resolution of
0\dasec1$\times$0\dasec2 and an rms noise of 0.104 mJy beam$^{-1}$.
The boxes (see Table 1) are regions where proper motion measurements
have been carried out with a $>3 \sigma$ level detection in at least
one direction except Box 21. (b) The H92$\alpha$ velocity distribution
with spectral and spatial resolutions of 14 \kms\ and
0\dasec75$\times$1\dasec2 presented in pseudo-color ({\it right}) with
contours of the $\lambda$2cm continuum image superposed.  The length
and direction of the vectors represent the transverse velocity in each
box.  The velocity and uncertainty of each vectors are presented in
columns 6 and 7 of Table 1.}}
\end{figure}

\begin{figure}
\plotone{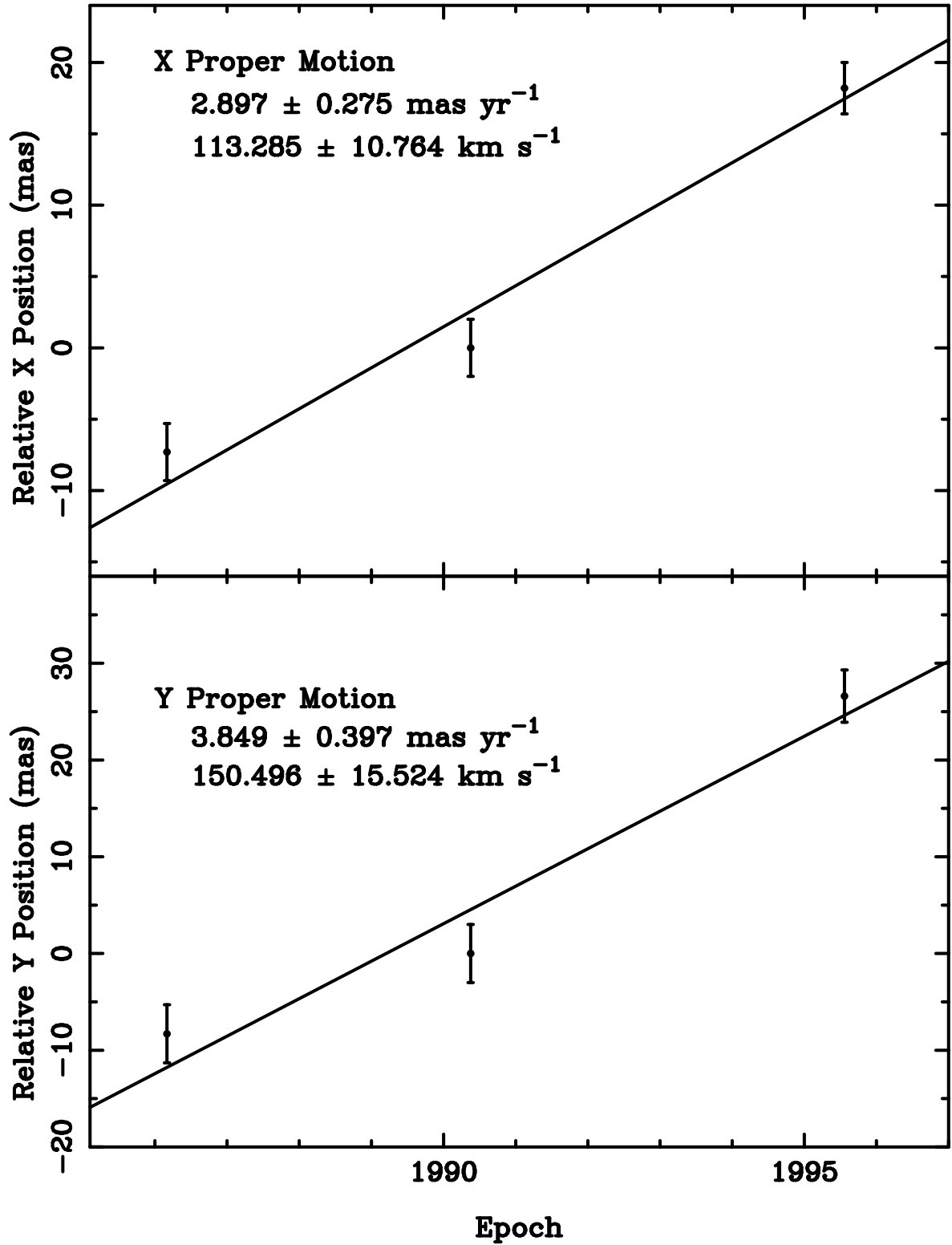}
\figcaption{The change in $\alpha$ and $\delta$ as a function of three
epochs for the region surrounding the IRS 13 cluster (Box 6 in Table
1).}
\end{figure}

\begin{figure}
\plotone{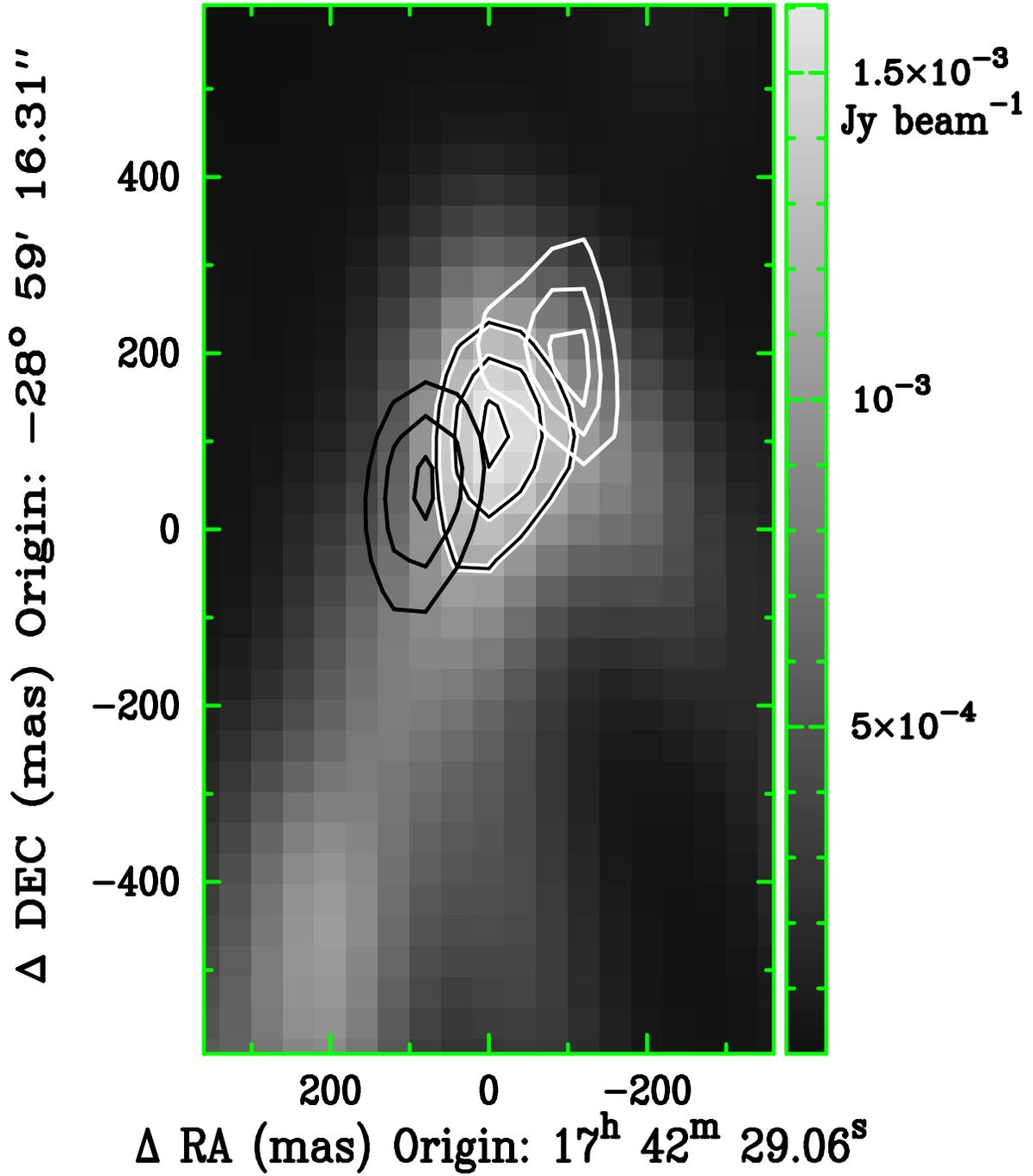}
\figcaption{The contours of the peak continuum emission from the
Bullet corresponding to three different epochs are superposed on the
grayscale image of the source (including the tail) from the 1990 epoch
observations.}
\end{figure}

\begin{figure}
\plotone{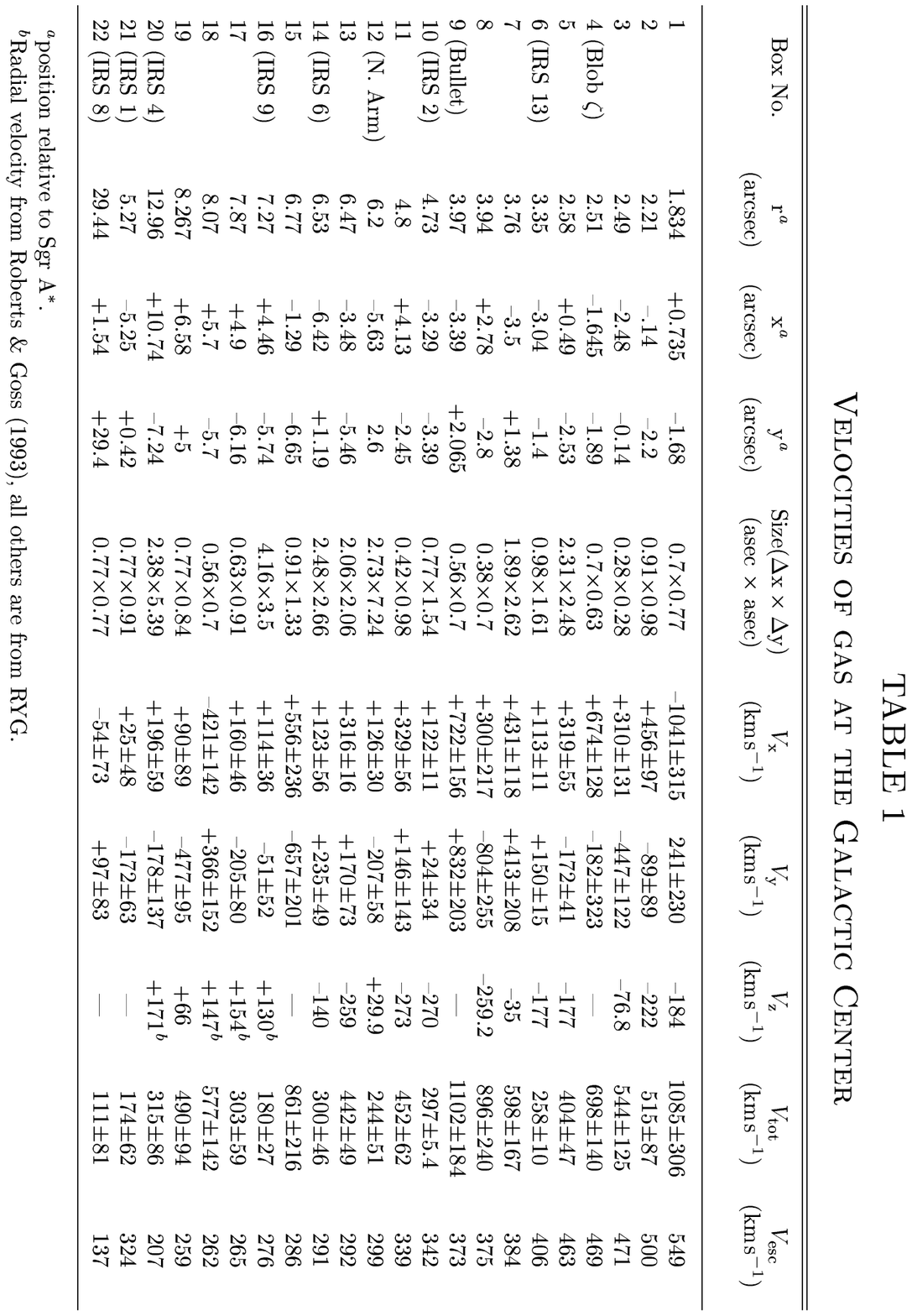}
\end{figure}

\end{document}